\documentstyle[preprint,aps]{revtex}
\begin{document}
\draft

\title{Self-Consistent Model of Annihilation-Diffusion Reaction\\
with Long-Range Interactions}
\author{Valeriy V. Ginzburg, Leo Radzihovsky and Noel A. Clark}
\address{Department of Physics, University of Colorado,
Boulder CO 80309-0390 USA}
\date{}
\maketitle

\begin{abstract}We introduce coarse-grained hydrodynamic equations of motion for 
diffusion-annihilation system with a power-law long-range interaction.
By taking into account fluctuations of the conserved order parameter -
charge density - we derive an analytically solvable approximation for
the nonconserved order parameter - total particle density.  Asymptotic
solutions are obtained for the case of random Gaussian initial
conditions and for system dimensionality $d \geq 2$.  Large-t,
intermediate-t and small-t asymptotics were calculated and compared
with existing scaling theories, exact results and simulation data.
\end{abstract}
\pacs{PACS: 05.60.+w, 82.40.-g, 61.30Jf}

\narrowtext
\section {Introduction}

In recent years, the annihilation-diffusion problem has generated
significant interest, both theoretical and experimental. The
annihilation-diffusion problem usually corresponds to that of the
kinetics of particle density decay in the annihilation reaction $A + A
\longrightarrow \emptyset$ (one-species annihilation) and $A + B
\longrightarrow \emptyset$ (two-species annihilation). In the latter
case, in most physical systems that one is interested in, in addition
to the thermal diffusion and kinetic annihilation, the particles A and
B are charged and interact via a power-law long-range interaction
(LRI), usually of Coulomb type, although here we will consider a more
general LRI. Such physically important interaction can clearly
strongly influence the annihilation dynamics by introducing an
additional time scale in the annihilation process, and can lead to a
new mechanism for slow dynamics. The two-species annihilation reaction
with LRI can be studied in a variety of experiments.  A thermal quench
of a freely-suspended liquid crystal film from smectic-A to smectic-C
phase \cite{Pindak,Muzny} is one experimental system where this
annihilation process has been studied in great detail. In such
experiments, immediately after quench, the singularities of the
smectic director (2d vector) field appear as positive and negative
vortices interacting (due to elastic forces) via a logarithmic
potential. As time elapses, vortices of opposite sign slowly
annihilate, exhibiting complex dynamics that is clearly and strongly
influenced by temperature, initial particle distribution, and LRI
that leads to an attraction between annihilating
partners. Similar annihilation problems also appear in turbulent flow,
superconductivity, spinodal decomposition, and many other condensed
matter systems. The annihilation process is also relevant to
coarsening of topological defects produced by symmetry breaking field
in particle physics models, after an early temperature quench due to the
fast initial universe expansion, a process that is thought in part to
determine the large scale structure of today's
universe.\cite{Chuang}

It is well known that in classical chemical kinetics, density decay for
both one-species and two-species annihilation is described by the
kinetic rate-equation (see, e.g., \cite{Entelis} ):
\begin{equation}
	\frac{d\rho}{dt} = - {\cal K} \rho^{2},	
	\label{eq:kinrate}
\end{equation}
with the large-$t$ asymptotics given by:
\begin{equation}
	\rho(t) \simeq ({\cal K} t)^{-1}\;,		
	\label{eq:mfsoln}
\end{equation}
where $\cal K$ is a reaction constant. Equation (\ref{eq:kinrate})
completely neglects all spatial fluctuations and correlations of the
particle density.  Although the dynamics described by
equations~(\ref{eq:kinrate}) - (\ref{eq:mfsoln}) is correct above an
upper critical space dimension $d_{uc}$, for $d\leq d_{uc}$, both diffusion and 
fluctuations play an important role, substantially slowing down the
density decay. It was indeed shown~\cite{Wilczek}, that for the
one-species annihilation without LRI, the $d_{uc}=2$. That is, for
systems of dimensionality less than $2$, the density decay is given by:
\begin{equation}
	\rho(t) \simeq \rho_0 (D \rho_{0}^{(2/d)} t)^{-d/2},	
	\label{eq:difsolnAA}
\end{equation}
where $D$ is diffusion constant and $\rho_{0}$ is the initial particle
density. Equation~(\ref{eq:difsolnAA}) can be understood either by invoking a
single length scaling arguments~\cite{Wilczek} or by using a more elaborate
Smoluchowski approach~\cite{Smoluchowski,Oshanin}. For the two-species case,
it was shown
(Refs.\cite{Wilczek}, \cite{Burlatsky,Zeldovich,Kang,Lebowitz})
that the upper critical dimension $d_{uc}=4$, and that for $d<4$,
the large-$t$ asymptotics is:
\begin{equation}
	\rho(t) \simeq (\rho_0)^{1/2} (D t)^{-d/4}.	
	\label{eq:difsolnAB}
\end{equation}
The decay law~(\ref{eq:difsolnAB}) was confirmed in several numerical
simulations of one-, two-, and three-dimensional
systems~\cite{Wilczek,Leyvraz,Jang}. It is important to emphasize that
in order to observe such power-law decay, it is necessary initially to
have an equal number of positive and negative charges, distributed at
random. If initial numbers of positive and negative charges are
different, one should observe an exponential density decay to the
non-vacuum equilibrium (see, e.g., Sokolov et al.~\cite{Sokolov1}).  If the
system is well-stirred, i.e., long-wavelength fluctuations of charge
density are suppressed, the decay law would also be different. We will
not consider such cases in this paper.

To better study the role of correlations in reactions without LRI, description
in terms of secondary quantization operators of creation and annihilation was 
proposed.~\cite{Doi,Pronin}
Further developing this approach, Peliti~\cite{Peliti} proposed an "exact"
(valid to all orders in perturbation expansion) renormalization-group
theory for the
one-species annihilation, and Lee and Cardy~\cite{Cardy} suggested a
similar approach to the two-species annihilation, both based on the
rigorous master equation converted into a field-theoretic
formulation. In the latter case, it was shown that the decay
law~(\ref{eq:difsolnAB}) is an asymptotic one for large $t$ and small
$\rho_{0}$. It was also rigorously proven that the assumptions used in the
derivation of equation~(\ref{eq:difsolnAB}) are correct only for $d
\geq 2$, while for $d < 2$, a more elaborate renormalization group
procedure which includes an effective noise with nontrivial correlations
should be carried out.

The addition of the LRI further complicates the picture. Equations of
motion become strongly coupled via the nonlocal interaction,
and only approximate solutions of these equations can be found. Up to
date, only Coulombic~\cite{Coulomb} systems in two dimensions ($d=2, n=1$, the case
corresponding to particles interacting via a logarithmic potential on
a two-dimensional substrate or film) were studied
numerically~\cite{Jang,DeGrand,Mondello,Huber}, and results appear to
be inconclusive.  In these simulations, particle density
exhibited a power-law decay $\rho(t)\sim t^{-\nu}$ with the exponent $\nu$
varying from $0.79 \pm 0.04$~\cite{Huber} to $0.85 \pm
0.05$~\cite{Jang}. The authors proposed a scaling
theory~\cite{Jang,Ginzburg} suggesting that for a Coulombic ($n = d - 1$)
two-dimensional diffusion-annihilation system, annihilation exponent
is equal to $0.85$ - close to but less than the mean-field exponent
$\nu=1$. Oshanin et al.~\cite{BOO} suggested that the exponent should be
exactly 1; Ispolatov and Krapivsky~\cite{Krapivsky} came to the same 
conclusion using an "inpenetrable domain" scaling theory. In their theoretical
and computational study of the defect annihilation in two-dimensional XY-model,
Yurke et al.~\cite{Huse} argued that annihilation exponent is 1, with
logarithmic correction due to the logarithmic dependence of the mobility on
the defect size. Thus, the problem of annihilation behavior for Coulombic
system in two dimensions has not been completely resolved, although it seems
rather plausible that the final asymptotics is governed by a classical exponent 1,
with possible logarithmic correction. It seems to be certain, however, that
annihilation in three-dimensional Coulombic system has a mean-field type final
asymptotics with exponent 1.

Finally, we should point out that until recently, there has been no discussion
of LRI other than Coulombic. Recently. several scaling theories have been
suggested to analyze arbitrary power-law interaction (more short-ranged than
Coulombic)~\cite{Ginzburg,Krapivsky,BGC}.Such problems may arise, e.g., in describing interactions between vacancies and interstitials in a two- or three-dimensional crystal.

In this paper, we propose a self-consistent theory based on
coarse-grained hydrodynamic equations of motion 
for particle number density and
charge density fields. This theory allows us to systematically
calculate the dependence of the density decay law on initial
conditions and to investigate the role of LRI. It can be
shown\cite{Ginz96} that the self-consistent approximation that we
employ here is equivalent to a resummation of an infinite class of
Feynman graphs which take into account the conserved charge density
fluctuations but ignore the less important nonconserved number density
fluctuations. This approximation can be further systematically
improved by the use of perturbation theory and the renormalization
group analysis that will be a subject of future
publication.\cite{Ginz96} In the limit of weak long-range
interactions, this approach agrees well with the known theoretical
results (see Refs. \cite{Wilczek,Cardy,Sokolov,Kuzovkov,Lindenberg})
for the two-species annihilation $A + B \longrightarrow \emptyset$,
thereby further clarifying the underlying assumptions that led to
these results.

The paper is organized as follows: in section~\ref{sec:eqns} we define
all variables and present the equations of motion on which our
analysis and results will be based. In section~\ref{sec:solns} the
self-consistent approximation is described and the asymptotic
solutions are obtained and analyzed. Finally, in
section~\ref{sec:concl} we analyze the resulting phase diagram and
discuss it in the context of the previously obtained results for
various diffusion-annihilation systems, which are selected points on
our phase diagram.

\section{Equations of motion}
\label{sec:eqns}

Let us consider a system consisting of two kinds of particles, A and B,
with A having a positive charge $+q$ and B having a negative charge
$-q$. We label their, time- and position-dependent concentrations as
$n_{1}({\bf r},t)$ and $n_{2}({\bf r},t)$, respectively, and impose
the condition that $<n_{1}({\bf r},t=0)>\ =\ <n_{2}({\bf r},t=0)>\ =\
n_{0}$.  The equation of motion for the densities is based on 
the generalized law of mass conservation, violated by the annihilation
process, 
\begin{equation}
	\frac{\partial n_{i}({\bf r},t)}{\partial t} + 
        \bbox{\nabla}\cdot{\bf J_{i}} = - {\cal K} 
	n_{1}({\bf r},t) n_{2}({\bf r},t),	
	\label{eq:evol}
\end{equation}
where the mass current is given by
\begin{equation}
	{\bf J_{i}} = - D\bbox{\nabla} n_{i}({\bf r},t) - 
         \mu q_{i} n_{i}({\bf r},t) \bbox{\nabla} V({\bf r},t),
		\label{eq:flux}
\end{equation}
and 
\begin{equation}
         V({\bf r},t) = q \int d^{d}{\bf r'} \frac{n_{1}({\bf r'},t)-
         n_{2}({\bf r'},t)}{|{\bf r'}-{\bf r}|^{n-1}}
		\label{eq:poten}
\end{equation}
is the electrostatic long-range potential at a point {\bf r} at time
$t$ due to local charge fluctuations away from neutrality; $n$ is the
power exponent of the long-range force, $\mu$ is particle mobility
taken to be a constant, and $q$ is particle
charge. Equations~(\ref{eq:evol}) - (\ref{eq:poten}) should be solved
in conjunction with initial conditions for $n_{1}({\bf r},t=0)$ and
$n_{2}({\bf r},t=0)$. Since here we are interested in statistical
averages, rather than in a dynamic solution for a given system with
specific initial conditions, we will focus on the density correlation
functions, with averages over random $t=0$ initial conditions.

Equations~(\ref{eq:evol}) - (\ref{eq:poten}) represent the coarse-grained continuum
limit for the "real" equations of motion; all variables in these equations
are averaged over "elementary volume" $n_{0}^{-1}$. Thus, only long-wavelength
modes are actually described by these equations, and, therefore, only 
intermediate-time and large-time regimes can be analyzed.

It is more natural and convenient to describe the system in terms of
the particle number and charge densities. We denote the former one as
$\rho({\bf r}, t)$, and the latter one as $f({\bf r}, t)$ and relate
them to densities $n_{1}$ and $n_{2}$ as follows:
\begin{equation}
	\rho({\bf r}, t) = 
         \frac{1}{2} (n_{1}({\bf r},t)+n_{2}({\bf r},t))\;,	
	\label{eq:rhodef}
\end{equation}
\begin{equation}
	f({\bf r}, t) = 
        \frac{1}{2} (n_{1}({\bf r},t)-n_{2}({\bf r},t)).\;.	
	\label{eq:fdef}
\end{equation}
Rewriting equations~(\ref{eq:evol}) - (\ref{eq:poten}) using
densities $f$ and $\rho$, we obtain:
\begin{eqnarray}
\label{eq:rhoevol}
	\frac{\partial \rho({\bf r},t)}{\partial t} - 
         D \nabla^{2} \rho({\bf r},t) &=& - {\cal K} 
	(\rho^{2}({\bf r},t) - f^{2}({\bf r},t)) - 
         Q \nabla (f({\bf r}, t) \nabla \int d^{d}{\bf r'} 
	\frac{f({\bf r'},t)}{|{\bf r'}-{\bf r}|^{n-1}}\;,\\
	\frac{\partial f({\bf r},t)}{\partial t} - D \nabla^{2} f({\bf
         r},t) &=& - Q \nabla (\rho({\bf r}, t) \nabla \int d^{d}{\bf
         r'} \frac{f({\bf r'},t)}{|{\bf r'}-{\bf r}|^{n-1}}\;.
\label{eq:fevol}
\end{eqnarray}
where $Q = \frac{1}{2} \mu q^{2}$. 

In the absence of LRI, equations~(\ref{eq:rhoevol}) - (\ref{eq:fevol})
are those analyzed in
Ref. \cite{Wilczek,Lebowitz,Cardy,Lindenberg}. However, the presence
of the additional long-range interactions, makes the analysis of their
asymptotic solutions rather nontrivial and, as we will show below, leads
to new dynamic regimes.

It is important to note that equations~(\ref{eq:rhoevol}) -
(\ref{eq:fevol}) do not contain noise terms in their right-hand
sides. It has been rigorously shown that such noise terms represent
important correlations and in some cases may even become predominant
in determining the asymptotic decay rate. In equations of motion
describing a near-equilibrium dynamics, powerful
fluctuation-dissipation theorem determines the form of noise
correlations. In contrast, in systems far from equilibrium, such as a
system of annihilating particles, it can be shown\cite{Cardy,Ginz96}
that the effective hydrodynamic equations of motion derived from the
fundametal master equations contain noise terms with very nontrivial 
correlations, of the form which could not be easily guessed a
priori. Lee and Cardy \cite{Cardy} proved that in a two-species
reaction without LRI, such noise leads only to the renormalization of
the reaction rate $\cal K$, but not to the change of the scaling
exponents, provided that space dimensionality $d > 2$. We have shown,
in a similar fashion \cite{Ginz96}, that for systems with LRI, the
noise has no effect on the asymptotic dynamics for $d > 2$, if the
renormalization of both $\cal K$ and $Q$ is implied. In addition,
since equations~(\ref{eq:rhoevol}) - (\ref{eq:fevol}) provide a
coarse-grained description on the length scale larger than the
interparticle spacing $\rho_{0}^{-1/d}$, the kinetic coefficients,
e.g. $\cal K$, are effective coefficients that incorporate finite
renormalization due to the correlations on short length scales.

In order to simplify further analysis, we divide each of the
equations~(\ref{eq:rhoevol}) -(\ref{eq:fevol}) by $\rho_{0}$ and
transform everything to dimensionless variables as follows:

\begin{math}
	 \rho \rightarrow \rho / \rho_{0}, \hspace{0.5in}  f \rightarrow f / \rho_{0},\hspace{0.5in} D \rightarrow D (\rho_{0})^{2/d},
\end{math}\\
\begin{math}
	 {\cal K} \rightarrow {\cal K}\rho_{0}, \hspace{0.5in} r \rightarrow r(\rho_{0})^{2/d}, \hspace{0.5in} Q \rightarrow Q(\rho_{0})^{\frac{n+1}{d}}.
\end{math}

It is important to notice that equation~(\ref{eq:fevol}) is linear
with respect to $f$, while equation~(\ref{eq:rhoevol}) is quadratic
with respect to $f$ (this points to the system's invariance with
respect to the simultaneous charge sign reversal for all
particles). In the next section, we will describe the self-consistent
approximation and its solutions.

\section{Self-Consistent Approximation and Solutions of Equations of Motion}
\label{sec:solns}

\subsection{Self-Consistent Approximation}

We now make an important approximation in order to further simplify
the analytical treatment of equations~(\ref{eq:rhoevol})-(\ref{eq:fevol}),
namely, we choose to ignore the fluctuations of the particle density
$\rho$ and concentrate on the fluctuations of the conserved charge
density $f$. This assumption is somewhat similar in spirit to the
approach of Glotzer and Coniglio~\cite{Glotzer} for the problem of
spinodal decomposition, and to the spherical approximation for the
Ising model in the limit of $N \longrightarrow \infty$. Unlike the
"classical" mean-field approach, however, the proposed approximation
does take into account exactly the charge density fluctuations, and is
expected, therefore, to describe at least some of the features of the
fluctuation-dominated kinetics.

The justification of the proposed assumption lies in a simple
observation that, while average particle density at any time is
nonzero, so that $<(\rho - <\rho>)^{2}>/<\rho>^{2}$ is finite and
likely to be small, average charge density is always zero, and
therefore in comparison the charge fluctuations $<(f - <f>)^{2}>$ are
large. Thus, we expect the former fluctuations to be less important
that the latter, and that we can approximate the particle number
density by its average (time-dependent) value in the equations of
motion without losing their important features. In a sense, our
approximation is a generalization of an argument used by Toussaint and
Wilczek~\cite{Wilczek}, in which they based their scaling decay law on
a suggestion that $<\rho> \approx \sqrt{<\rho^{2}>}$. It seems clear
that the approximation of ignoring the number density fluctuations
must breakdown at least below some upper-critical dimension $d_{uc}$,
since asymptotically $\rho$ vanishes. In this case our approximation
will be valid for $d>d_{uc}$ for all times and in systems below
$d_{uc}$ it will be a good approximation up to a crossover time beyond
which the asymptotics will be modified by the number density
fluctuations. Systematically taking into account these additional
fluctuations will be a subject of future work.\cite{Ginz96}

Taking into account the above approximation, we rewrite the
equation~(\ref{eq:fevol}) in Fourier representation, taking
$\rho(t)$ as a spatially-independent but time-dependent function:
\begin{equation}
	(\frac{\partial}{\partial t} + D k^{2} + 
        Q \rho(t) k^{2-\sigma}) f({\bf k}, t) = 0,
\label{eq:fkevol}
\end{equation}
where $\sigma = d + 1 - n$.

The equation~(\ref{eq:rhoevol}) in this self-consistent approximation
is rewritten as:
\begin{equation}
	\frac{d \rho}{d t} + {\cal K}\rho^{2} = {\cal K}\int \frac{d^{d} k}{(2 \pi)^{d}}
	<f({\bf k}, t) f({\bf -k}, t)>,
\label{eq:rhonewevol}
\end{equation}
where $<\ldots>$ denotes averaging over initial conditions.

These equations of motion have to be supplemented with initial
conditions. It is well known that the initial density distribution
plays an important role in determining the scaling decay law. Although the
self-consistent approximation employed here is well suited for a
comprehensive study of the influence of initial conditions on the
dynamics, here we limit our study to a single type of an initial
condition. Throughout this paper we will focus on the dynamics
initiated with a random Gaussian particle distribution, completely
characterized by
\begin{eqnarray}
	<f({\bf k},0)>&=&0\;,\label{eq:f1kinit}\\
	<f({\bf k_{1}}, 0)f({\bf k_{2}}, 0)>&=&\Delta\ (2
        \pi)^{d}\delta^{(d)}({\bf k_{1} + k_{2}})\;,
\label{eq:fkinit}
\end{eqnarray}
which constraints the system to charge neutrality at all times and for
simplicity we take the variance $\Delta=1$ ($=\rho_0^2$ in physical
units).  For most charged experimental systems a more relevant initial
conditions incorporate the suppression of long-wavelength fluctuations
in charge density, which can be modeled (within e.g. the
Debye-H\"{u}ckel approximation) by $\Delta(k)=\Delta_0
k^2/(k^2+k_s^2)$ in equation~(\ref{eq:fkinit}).

The diffusion-only (DO) case ($Q=0$) and Coulombic case ($n=d-1$) are
the simplest systems with a relatively clear, yet interesting asymptotic
behavior. All the intermediate interactions (arbitrary $n$ and $d$)
lead to a more complicated scaling behavior, with several regimes and
crossovers. We will devote a subsection to each of these three cases.

\subsection{Systems without Long-Range Interactions}

There are two ways of approaching the limit of "no long-range
interactions": by decreasing the force constant $Q$ to 0 or by
increasing the power exponent $n$ to infinity (interaction with an
effectively vanishing range). Obviously, these limits should give the
same answer. For simplicity, we will set $Q = 0$ and show that our
self-consistent approximation yields the well-known
results.\cite{Wilczek}-\cite{Lebowitz}
\begin{eqnarray}
	\rho(t) &\simeq& ({\cal K} t)^{-1}\;, \;\;\;\mbox{for}\;\; d > 4\;, \\
	\rho(t) &\simeq& (Dt)^{-d/4}\;, \;\;\;\mbox{for}\;\; d < 4\;.
\label{eq:twsoln}
\end{eqnarray}
For $Q=0$, the kinetic equation for $f$ reduces to a simple diffusion
equation, with the solution
\begin{equation}
	f({\bf k}, t) = f({\bf k}, 0)\;e^{-Dk^{2}t}\;.
\label{eq:twfk}
\end{equation}
Substituting this solution~(\ref{eq:twfk}) for $f({\bf k}, t)$ into
equation~ (\ref{eq:rhonewevol}), and taking into account the initial
condition~(\ref{eq:f1kinit}) - (\ref{eq:fkinit}), we obtain:
\begin{equation}
	\frac{d \rho}{d t} + {\cal K} \rho^{2} = 
        \frac{{\cal K}}{(1 + 2Dt)^{d/2}}.
\label{eq:twequation}
\end{equation}
An the exact solution expressible in terms of confluent hypergeometric
functions is possible.\cite{Cardy} It can also be easily shown that
equation~(\ref{eq:twsoln}) describes the asymptotic solution of
equation~(\ref{eq:twequation}). This is expected, since, as we argued
above, the approximations made by Toussaint and Wilczek~\cite{Wilczek}
are very similar to our self-consistent model.

The case of $n \longrightarrow \infty$ will be analyzed in
subsection~\ref{Intermediate Systems}, where it will be shown that for
all $n > 1 + d/2$, the decay law is asymptotically the same as for DO
systems.

\subsection{Coulombic Systems}

In Coulombic systems, the long-range interaction is the strongest
possible that one can achieve without making the system
thermodynamically unstable (systems with interactions stronger than
Coulombic have infinite pressure and chemical potential even if their
total charge is zero).  Because of this, one would expect the particle
density decay for Coulombic systems to be very close or equal 
to the mean-field law $\rho(t) \approx ({\cal K} t)^{-1}$.~\cite{Ohtsuki} 
As we describe below, the self-consistent approximation predicts the
decay exponent $\nu=1$
consistent with this expectation and with some simulations reported in
the literature.\cite{Huse} However, it can be shown\cite{Ginz96} that
correction to this mean-field like decay law can arise from the number
density fluctuations and noise for $d\leq 2$, both of which have been
neglected in the self-consistent theory presented here.

Equation for the evolution of charge density $f$ for the Coulombic systems
in the self-consistent approximation can be exactly solved to yield:
\begin{equation}
	f({\bf k}, t) = f({\bf k}, 0) \exp[-Dk^{2}t - 
        Q \int_{0}^{t} \rho(\tau) d\tau]\;.
	\label{eq:coulfksoln}
\end{equation}
Using this solution~(\ref{eq:coulfksoln}) and the initial
condition~(\ref{eq:f1kinit}) - (\ref{eq:fkinit}), $\rho(t)$ can be
easily shown to satisfy the following differential equation:
\begin{equation}
	\frac{d \rho}{d t} + {\cal K}\rho^{2} = {\cal K}
	\exp(-2Q\int_{0}^{t} \rho(\tau) d \tau) 
	\int \frac{d^{d} k}{(2 \pi)^{d}} \exp(-2Dk^{2}t),   
\label{eq:coulrhoevol}
\end{equation}
In order to find the asymptotic solutions, we introduce a new
variable:
\begin{equation}
	\Theta = \int_{0}^{t} \rho(\tau) d\tau.
\label{eq:Theta}
\end{equation}
Equation~(\ref{eq:coulrhoevol}) then transforms to:
\begin{equation}
	\frac{d^{2} \Theta}{dt^{2}} + {\cal K}\left(\frac{d
	\Theta}{dt}\right)^{2} = \exp(-2Q\Theta) \frac{{\cal K}}{(1 +
	2Dt)^{d/2}}.
\label{eq:Thetaevol}
\end{equation}
Let us find the "critical" dimension $d_{uc}$, above which the
mean-field behavior is manifested. The mean-field solution for
$\Theta$ is given by
\begin{equation}
	\Theta(t) = \frac{1}{\cal K} \ln (1 + {\cal K}t) + \ldots,
\label{eq:Thetamfsoln}
\end{equation}
where $\ldots$ corresponds to subdominant constant terms and terms
decreasing with time. By counting powers of $t$ in the right-hand side and
the left-hand side of equation~(\ref{eq:Thetaevol}), we obtain:

power(LHS) = $-2$;

power(RHS) = $-\frac{d}{2} -2\frac{Q}{\cal K}$.\\
Obviously, for the mean-field solution to be valid asymptotically, the
power(LHS) should be larger than the power(RHS), which happens for
systems with dimensionality larger than the critical dimension,
\begin{equation}
	d > d_{uc} = 4(1 - \frac{Q}{\cal K})\;.
\label{eq:coulDcr}
\end{equation}
If $Q \geq {\cal K}$, the asymptotic kinetics is determined by the
slower process, which is the annihilation, with possibly interaction
renormalized ${\cal K}$ (implicitely assumed here), and the Coulomb
interaction and diffusion are asymptotically irrelevant. For $Q
< {\cal K}$, the Coulomb interaction has an interesting effect of
continuously lowering the upper-critical dimension from $4$ (for
$Q/{\cal K}=0$, in which case for $d<4$ the diffusion dominates giving
$\nu=d/4$) down to $d_{uc}$ given above.

In order to analyze the kinetics when space dimensionality $d$ is
below $d_{uc}$, we employ the "steady-state" approximation, which
suggests that at long times the time derivative in the LHS of
equation~(\ref{eq:coulrhoevol}) is the smallest of the three terms. In
this case, the equation of motion can be written as:
\begin{eqnarray}
	\frac{d \Theta}{dt}
       &=& \exp(-Q \Theta) (1 + 2Dt)^{-d/4}\label{eq:Theta2evol}\;,\\
    \Theta(0) &=& 0\;.
\end{eqnarray}
An exact solution of this equation is:
\begin{equation}
	\Theta = \frac{1}{Q} \ln [1 + \frac{Q}{2 D (1 - d/4)}
	( \{1 + 2Dt\}^{1-d/4} -1 )]\;,
\label{eq:Theta2soln}
\end{equation}
and 
\begin{equation}
	\rho = \dot {\Theta}.
\end{equation}
It can be easily shown that for large $t$, the asymptotic solution
for the particle density decay ($d<d_{uc}\leq 4$) is:
\begin{equation}
	\rho \simeq \frac{1 - d/4}{Q t}\;,
\label{eq:coulrholarget}
\end{equation}
predicting the asymptotic decay exponent $\nu=1$ for Coulomb systems,
as in the mean-field regime, although with $Q$- rather than $\cal
K$-determined amplitude. This large-$t$ limit is achieved when:
\begin{equation}
	t > t_{L} = 
\frac{1}{2 D} \left [ \frac{2 D (1 - d/4)}{Q} \right ]^{{4}/(4-d)}\;,
\label{eq:coullarget}
\end{equation}
and it can be easily seen that in the limit of $Q \rightarrow 0$
(vanishing interactions) the transition time $t_L$ to this region becomes
infinite, i.e. this time is never reached.

If diffusion is faster than the deterministic Coulomb
interaction-driven relaxation, i.e., $D > Q$, then for times less than
$t_{L}$, the annihilation is governed by the intermediate asymptotics:
\begin{equation}
	\Theta = \frac{1}{(1 - d/4) D} (2 D t)^{1 - d/4},
\label{eq:coulintermed_t}
\end{equation}
so the particle density is described by the Toussaint-Wilczek solution
up to the crossover time $t_L$:
\begin{equation}
	\rho = \dot {\Theta} \simeq (D t)^{-d/4}
\label{eq:coulrhointermed_t}
\end{equation}
This intermediate asymptotics, which exists only for $d < 4$, reflects
early times diffusion-dominated decay, with the slower deterministic
Coulomb interaction driven classical $t^{-1}$ decay appearing only at
times later than $t_L$. In contrast, for $D < Q$ or if $d > 4$, there
is no extended intermediate regime and one should see a quick
transition to a classical decay law.  Although within the
self-consistent approximation the asymptotic $1/t$ decay is not
affected by the choice of reasonable initial conditions, the
intermediate diffusion-dominated decay is certainly affected by our
choice of random Gaussian uncorrelated initial conditions given in
equation~(\ref{eq:fkinit}). For instance if the screened Debye-H\"{u}ckel
initial conditions are used with $\Delta(k)=\Delta_0 k^2/(k^2+k_s^2)$,
then this $\Delta(k)$ will appear as a multiplicative kernel under the
$k$ integral in equation~(\ref{eq:coulrhoevol}). It will then modify the
intermediate decay exponent from $d/4$ to $\nu=(d+2)/4$ (and, for $d > 2$,
eliminating this intermediate region altogether), without
modifying the asymptotic decay of equation~(\ref{eq:coulrholarget}).

To sum up, we find that within the self-consistent approximation,
Coulombic systems $(n = d - 1)$ asymptotically exhibit the $t^{-1}$
density decay, consistent with several scaling arguments and
simulations.\cite{BOO,Krapivsky,Huse,Ohtsuki}

\subsection{Intermediate systems}
\label{Intermediate Systems}

Let us now consider the general case of long-range interactions with a
power-law $d-1 < n < \infty$ that is of shorter range (weaker) than
the Coulomb interaction considered in the previous section.
Equations~(\ref{eq:fkevol}) - (\ref{eq:rhonewevol}) can be solved to
yield:
\begin{eqnarray}
	f({\bf k}, t) &=& f({\bf k}, 0) \exp [- Dk^{2}t - Qk^{2-\sigma}
	\int_{0}^{t} \rho(\tau) d \tau]\;,\label{eq:intfksoln}\\
	\frac{d \rho}{d t} + {\cal K}\rho^{2} &=& 
	{\cal K}\int \frac{d^{d} k}{(2 \pi)^{d}} 
	\exp [-2Dk^{2}t - 2Qk^{2-\sigma} \int_{0}^{t} \rho(\tau) d \tau]\;,
\label{eq:intrhoevol}
\end{eqnarray}
where $\sigma = d - n + 1$.

Equation~(\ref{eq:intrhoevol}) is significantly more complicated than
its analogs for either Coulombic or non-interacting
cases. Nevertheless, it is possible to find its power-law asymptotic
solutions.  Using an asymptotic analysis analogous to that described in
the previous subsection, we find several kinetic regimes depending on the
values of $d$ and $n$. These regimes depend crucially on the charge
density relaxation mechanism, i.e., whether the LRI or diffusion
determines the relaxation rate of $f({\bf k},t)$ at late times.  In
order to analyze the asymptotic behavior of the system, we again
introduce the integrated density $\Theta (t)$ as defined in equation
(\ref{eq:Theta}). We also assume a power-law for the density and, for
$d < 4$, neglect the term $d\rho/d t$. In this case, the
equations of motion are:
\begin{eqnarray}
	\frac{d \Theta}{d t} &=& \sqrt{ \int \frac{d^{d} k}{(2 \pi)^{d}}
	\exp [-2Dk^{2}t - 2Qk^{2-\sigma} \Theta (t) ]}\;,\label{eq:ims1}\\
	\rho (t) &=& \frac{d \Theta}{d t}\;.
\end{eqnarray}
Depending on $\sigma$ and $d$, either the first or second term in the
exponential in equation (\ref{eq:ims1}) dominates for
large $t$, corresponding to either diffusive or superdiffusive
relaxation. We first assume that diffusive relaxation is prevalent and
determine the conditions when it is true. In the case of diffusive
relaxation mechanism and at large $t$, equation~(\ref{eq:ims1}) can be
simplified to yield:
\begin{equation}
	\frac{d \Theta}{d t} = \sqrt{k_{d} (D t)^{-d/2}
	\int_{0}^{\infty} x^{d/2 - 1} dx \exp [-x - 
        \frac{2Q\Theta(t)}{(2Dt)^{1 - \sigma / 2}} x^{1 - \sigma /2} ]}\;,
\label{eq:ims2}
\end{equation}
where $k_{d}$ is a dimensionless constant absorbing integration over
angular variables, and the integral becomes time-independent for large
$t$ if the second term in the exponential vanishes with time. If we
assume the power-law dependence for $\rho$ and $\Theta$:
\begin{eqnarray}
	\rho (t) &\propto& t^{- \nu}\;,\\
	\Theta (t) &\propto& t^{1 - \nu}\;,
\end{eqnarray}
then it follows from equation~(\ref{eq:ims2}) that for the predominantly 
diffusive system $\nu = d/4$, as expected. Thus, in order for this solution
to be self-consistent, we must require that:
\begin{eqnarray}
	1 - \nu &<& 1 - \sigma /2\;,\\
	\sigma &>& d/2\;,
\end{eqnarray}
and, from the definition of $\sigma$, we determine the region where
the relaxation and density decay are diffusion-limited:
\begin{eqnarray}
	n &>& 1 + d/2\;,\\
	d &<& 4\;.
\end{eqnarray}
This region is marked FD (fluctuation-dominated) in Figure 1. In it,
the LRI are irrelevant at large $t$, although they may influence the
density decay kinetics for intermediate $t$. The asymptotic decay law
in the FD region is given by:
\begin{equation}
	\rho \simeq (D t)^{-d/4}
\label{eq:fddecay}
\end{equation}
In the region referred to as the IR (intermediate region) and which
lies below FD in Figure 1 ($d - 1 \leq n \leq 1 + d/2$) the LRI are
strong and dominate the diffusion at large $t$. To investigate the
asymptotics of the decay in this region, we rewrite
equation~(\ref{eq:ims1}) in the following way:
\begin{equation}
	A (1 - \nu) t^{- \nu} = \sqrt{k_{d} (Q A t^{1 - \nu})^{-d/(2 - \sigma)}
	\int_{0}^{\infty} x^{(d/(2 - \sigma))-1} dx 
\exp{[-x - \frac{2 D t}{(2 Q A t^{1 - \nu})^{2/(2 - \sigma)}} x^{2/(2 - \sigma)}}]}.
\label{eq:ims4}
\end{equation}
using $\Theta = A t^{1 - \nu}$. By solving equation~(\ref{eq:ims4})
approximately we find two asymptotics in this region:
\begin{equation}
	\rho(t) \simeq (Dt)^{-d/4}
\end{equation}
valid at intermediate times, and
\begin{equation}
	\rho(t) \simeq (Qt)^{-\nu},	
\end{equation}
where 
\begin{equation}
	\nu = \frac{d}{4 + d - 2 \sigma} = \frac{d}{2 - d + 2n} 
\label{eq:nuir}
\end{equation}
for asymptotically large times.

The crossover time $t_{c}$ from the diffusion-dominated decay to the
LRI-dominated decay is:
\begin{equation}
	t_{c} \approx D^{\frac{2-d+2n}{2+d-n}} Q^{- \frac{2}{1+d/2-n}}.
\end{equation}
Thus, in this region (marked IR - intermediate region - in Figure 1)
the LRI accelerates the relaxation of the initial density fluctuations and
thereby speed up the annihilation. If $Q < D$, the diffusive
relaxation and the ($d/4$)-law may be observed for the intermediate
$t$ before transition to the superdiffusive relaxation and faster
decay takes place for $t>t_c$.

If $d > 4$ (region MF - mean-field - in Figure 1), spatial
fluctuations become irrelevant and the classical kinetic-rate equation
becomes asymptotically correct, so the decay law in this region is
given by:
\begin{equation}
	\rho(t) \simeq ({\cal K}t)^{-1},
\end{equation}
as previously discussed.

\section{Summary and Conclusions}
\label{sec:concl}

In the presented work we derive approximate kinetic equations for the 
annihilation-diffusion process with long-range forces. To analyze the
asymptotic decay law for systems with $d > 2$, we proposed a new
self-consistent method of calculating the average particle density as
function of time.  Since the total particle density is a nonconserved
order parameter with positive average at all times, we argued that its
fluctuations are less important in determining dynamics of
annihilation than that of a conserved order parameter - the charge
density. This approximation self-consistently decouples two kinetic
equations and make it possible to find the asymptotic solutions.

In the limit of weak long-range interaction (either via taking $n
\rightarrow \infty$ or $Q \rightarrow 0$), self-consistent equations
of motion are reduced to those of Toussaint and
Wilczek~\cite{Wilczek}.

For Coulombic systems in more than two dimensions, our model yields
the mean-field exponent $\nu=1$, yet the role of segregation (i.e.,
charge density fluctuations) is important and cannot be simply left
out. Ispolatov and Krapivsky~\cite{Krapivsky} proposed the
"unpenetrable domain" scaling concept, which in Coulombic case results
in decay exponent $1$ independently of space dimensionality.  Both
their model and our self-consistent approximation neglect possible
fluctuation modes due to the spatial variation of particle density, as
well as noise, which can lead to some slowing down of the reaction
kinetics, as indicated by simulations. Elucidation of such modes and
their role should require detailed account of noise and possibly use
of renormalization group analysis when $d = 2$, since it is the
critical dimension for the annihilation-diffusion problem.  Since
neglecting of noise in this problem appears to be justified for
long-time asymptotics for systems with $d > 2$,\cite{Ginz96} it is
possible that the exponent $\nu=1$ for $d>2$ Coulombic systems is
exact, even though there is no experimental evidence to support this
conclusion.

The analysis of the self-consistent equations of
motion~(\ref{eq:fkevol}) - (\ref{eq:rhonewevol}) suggests that there
is a new region in the (n,d) phase diagram (the region labeled IR in
Figure 1), in which the large-t asymptotics of the density decay is
determined by long-range forces. The annihilation initally depletes
positively charged region of negative particles and vice versa, and
then the decay rate is determined by the speed of particle drift from
such regions. Again, our large-t asymptotics here agrees with the
"unpenetrable domain" theory of Ispolatov and Krapivsky, although
their model predicts different boundaries of the IR region in the
(n,d) phase space ($n = 1 + d/2$ is the upper boundary of the IR
region in the self-consistent model, and $n = d$ is the upper boundary
of the IR region in the "unpenetrable domain model"; the lower
boundary in both theories is the Coulombic line $n = d - 1$). The
self-consistent theory also predicts a crossover from
diffusion-dominated decay to the LRI-dominated decay at large times for
the systems in this region.

Because the self-consistent model is a semi-mean-field approximation
(it completely neglects particle density fluctuations and only takes
into account the concerved charge density fluctuations), it should be
considered only as a first step. A systematic perturbative 
analysis of equations~(\ref{eq:rhoevol}) - (\ref{eq:fevol}) around our
self-consistent solution is needed to assess the role of left out
number density fluctuations and noise.\cite{Ginz96}

The proposed self-consistent model, its somewhat uncontrolled
approximations notwithstanding, represents an important tool in
qualitative analysis of dynamic processes in two-component systems
with one conserved and one non-conserved variable. It predicts new
annihilation behavior (IR regime) and new crossovers between
diffusion-driven and LRI-driven decay regions, reproduces all known
results for the annihilation problem in special limits, and can be
used to systematically study the role of initial conditions in such
processes.

\vspace{0.5in} 

{\bf Acknowledgment}. Leo Radzihovsky was supported by the 
National Science Foundation CAREER award, through Grant DMR-9625111.
We thank Dr. B. Lee and Dr. G. Oshanin for helpful
comments and suggestions.

\begin{figure}Figure 1. Phase diagram of the annihilation-diffusion 
reaction with long-range forces. FR - forbidden region (below
Coulombic line). IR - intermediate region - large-$t$ asymptotics is
determined by LRI. FD - fluctuation-dominated region - large-$t$
asymptotics is determined by diffusion and initial fluctuations. MF -
mean-field region - large-$t$ asymptotics is determined by the
kinetic-rate equation.
\end{figure}


\begin{thebibliography}{99}
\bibitem{Coulomb} By Coulombic, we mean such a LRI that its potential
obeys Poisson's equation in d-dimensional system. This transforms to
a relationship between system dimensionality d and LR force power 
exponent n: $n = d - 1$.
\end{thebibliography}
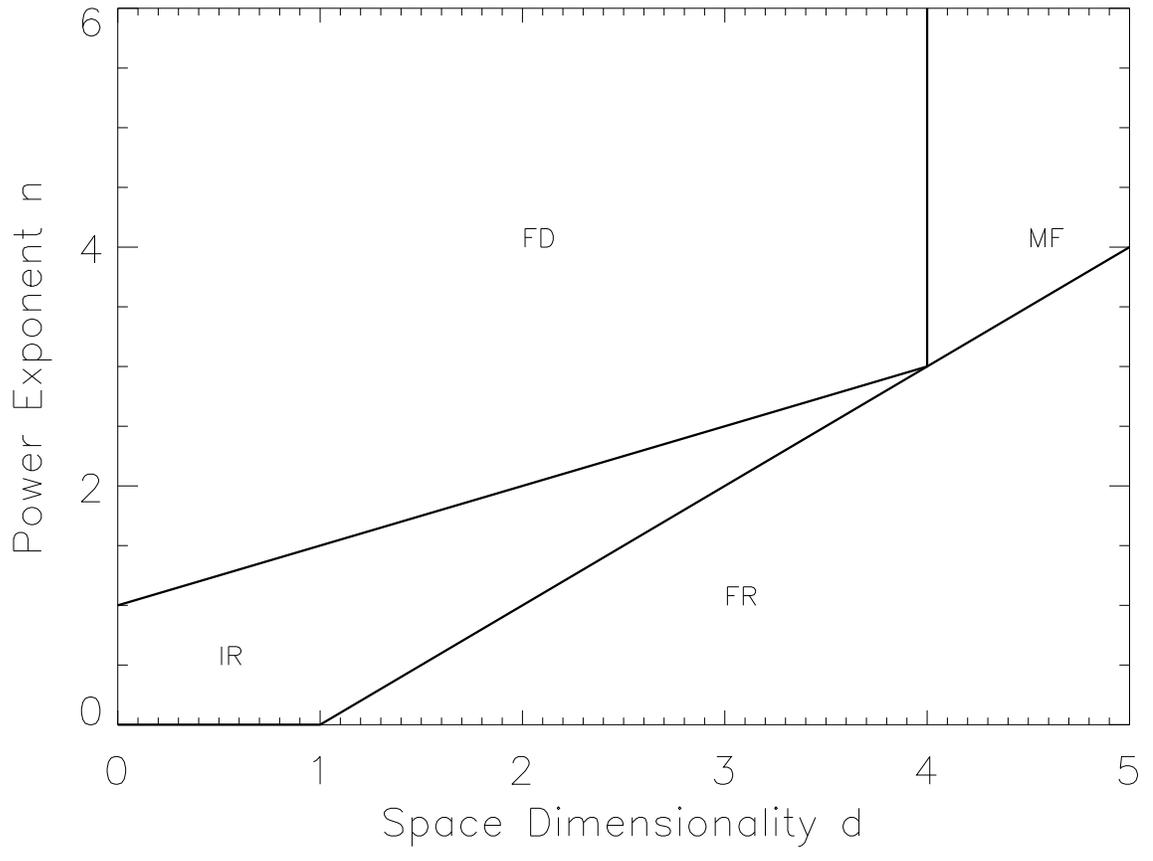
\end{document}